\documentclass{epl}

\title{Gravitational decoherence of planetary motions}
\shorttitle{Gravitational decoherence}
\author{S. Reynaud \inst{1} 
\thanks{E-mail: \email{reynaud@spectro.jussieu.fr}}  
\and P. A. Maia Neto \inst{2} 
\and A. Lambrecht \inst{1} 
\and M.-T. Jaekel \inst{3} }
\shortauthor{S. Reynaud \etal}
\institute{\inst{1}Laboratoire Kastler Brossel
\thanks{Laboratoire ENS, Universit\'e Pierre
et Marie Curie, CNRS}, Campus Jussieu case 74, 
F75252 Paris, France \\
\inst{2}Instituto de F\'{\i}sica, UFRJ, Caixa Postal 68528, 
21945-970 Rio de Janeiro, Brazil \\
\inst{3}Laboratoire de Physique Th\'{e}orique
\thanks{Laboratoire CNRS, ENS, Universit\'e Paris-Sud},
24 rue Lhomond, F75231 Paris, France}
\rec{November 2000}{}


\begin{document}

\maketitle

\begin{abstract}
We study the effect of the scattering of gravitational waves on 
planetary motions, say the motion of the Moon around the Earth.
Though this effect has a negligible influence on dissipation, 
it dominates fluctuations and the associated decoherence mechanism, 
due to the very high effective temperature of the  
background of gravitational waves in our galactic environment.
\end{abstract}

Decoherence plays an important role in the transition between microscopic
and macroscopic physics since it kills quantum coherences on a time scale 
which becomes extremely short for systems with a large degree of 
classicality \cite{Zeh70,Dekker77,Zurek81,Joos85}. The details of this 
transition depend on the coupling mechanisms between the system under 
consideration and its environment. A lot of different models have been 
considered theoretically and decoherence has been experimentally observed 
in mesoscopic systems, such as a few microwave photons in a high-Q cavity 
\cite{Brune96}, for which the decoherence time is neither too short nor 
too long.

For large macroscopic masses, say the Moon orbiting around the Earth,
decoherence is so efficient that it precludes the observation of quantum 
coherences. It remains however important from a conceptual point of view 
to study the dominant decoherence mechanisms and to obtain a reliable
estimation of the decoherence time scale. For motions in the solar system, 
decoherence is often attributed to the scattering of the electromagnetic 
fluctuations associated with solar radiation or cosmic microwave background. 
In this letter we show that the decoherence of planetary motions is not
dominated by electromagnetic processes but rather by the scattering of 
stochastic gravitational waves present in our galactic environment.

Gravitational fluctuations have already been proposed as a universal 
mechanism able to explain the transition from quantum to classical physics. 
If these fluctuations are characterized by length scales of the order of
the Planck length \cite{Wheeler57,DeWitt62}, microscopic and macroscopic 
regions may be delineated by comparing the Planck length 
$l_{\rm P}=\sqrt{\frac{\hbar G}{c^{3}}}$ and the Compton length 
$l_{\rm C}=\frac{\hbar }{mc}$ associated with typical quantum phenomena 
for a mass $m$ \cite{Karolyhazy66,Diosi89,Penrose96}. Remarkably, this 
leads to a frontier determined by the Planck mass 
$m_{\rm P}=\sqrt{\frac{\hbar c}{G}}$, i.e. the mass scale built on
the Planck constant $\hbar$, the velocity of light $c$ and the Newton
constant $G$ with a value $22\ \mu {\rm g}$ lying on the borderland 
between microscopic and macroscopic masses. However, this simple scaling 
argument is not by itself sufficient to reach any precise conclusion.

In order to compare gravitational and electrodynamical contributions
to decoherence, one has to discuss the corresponding fluctuation levels 
as well as their effects on the system, at the frequencies of interest 
for the latter. In this letter, we present a quantitative study of 
decoherence of planetary motions in the stochastic background of 
gravitational waves in our environment \cite{Hils90,Giazotto97}. We give 
a precise estimate of the associated gravitational decoherence rate and 
show that it largely dominates the electromagnetic contribution.

The interaction of macroscopic motions with gravitational fluctuations is
well known from the theory of gravitational wave emission and gravitational
wave detection \cite{Landau,Misner,Blanchet00}. At the limit of non 
relativistic velocities, the gravitational perturbation on a planetary
system may be represented as a tidal acceleration acting on each mass 
\begin{eqnarray}
&&x_{\rm i}^{\prime \prime} \left( t\right)  = -R_{\rm i0j0} \left( t\right) 
x_{\rm j}\left( t\right) \qquad 
R_{\rm i0j0} \left( t\right) = -\frac{1}{2} h_{\rm ij}^{\prime \prime}
\left( t\right)
\end{eqnarray}
The prime denotes a time derivative.
The tidal tensor $R_{\rm i0j0}$ is built up from components of the Riemann 
curvature tensor with the index ${\rm 0}$ representing time
components and indices ${\rm i,j}$ representing spatial components. 
$R_{\rm i0j0}$ has been written as the second order derivative of the 
metric perturbation $h_{\rm ij}$ evaluated at the center of mass of 
the planetary system in the transverse traceless (TT) gauge.
The interaction is equivalently described by a Lagrangian perturbation
coupling the quadrupole momentum of the system to the tidal tensor
\begin{eqnarray}
&&S^{\prime }\left( t\right) = \frac{1}{4} h_{\rm ij}^{\prime \prime}
\left( t\right) Q_{\rm ij}\left( t\right)  \qquad
Q_{\rm ij}\left( t\right)  = m \left( x_{\rm i}\left( t\right) 
x_{\rm j}\left( t\right) - \frac{\delta_{\rm ij}}3
x_{\rm k}\left( t\right) x_{\rm k}\left( t\right) \right)
\label{Lagrangian}
\end{eqnarray}
The quadrupole $Q_{\rm ij}$ is reduced to its traceless part 
with $\delta_{\rm ij}$ a spatial Kronecker symbol. 
$S^{\prime }$ is the time derivative of the action integral as it is 
perturbed along a given trajectory, for example a circular orbit
in the plane $x_{\rm 1}x_{\rm 2}$ 
\begin{eqnarray}
&&x_{\rm 1}\left( t\right) = \rho \cos \left( \Omega t+\theta \right) 
\qquad 
x_{\rm 2}\left( t\right) = \rho \sin \left( \Omega t+\theta \right)   
\qquad
\rho ^{3}\Omega ^{2} = GM
\label{Kepler}
\end{eqnarray}
For a system of two masses $m_{a}$ and $m_{b}$,
$m$ denotes the reduced mass $\frac{m_{a}m_{b}}{m_{a}+m_{b}}$,
$x_{\rm i}$ the relative position and $\rho$ the distance between
the two masses. The last relation in (\ref{Kepler}) is the Kepler law which
connects the radius $\rho$, the orbital frequency $\Omega$
and the total mass $M=m_{a}+m_{b}$. 

Our aim is to evaluate decoherence between two neighbouring motions
on the same circular orbit. These motions correspond to slightly 
different values of $\theta$ with the difference denoted by 
$\Delta \theta$. The differential perturbation between the two 
motions is thus described by a quantity $\Delta S^{\prime }$
written as in (\ref{Lagrangian}) with $Q_{\rm ij}$ replaced by
the difference $\Delta Q_{\rm ij}$ of the quadrupoles 
on the two motions. We write it as the product of a distance 
$\Delta x$ and a force $F$  
\begin{eqnarray}
&&\Delta S^{\prime }\left( t\right) =F\left( t\right) \Delta x\qquad
\Delta x=\rho \Delta \theta   \qquad
F\left( t\right) = \frac{1}{2} h_{\rm ij} ^{\prime \prime} \left( t\right) 
m x_{\rm i} \left( t\right) \frac{x_{\rm j}^{\prime } \left( t\right)} 
{\rho \Omega }
\end{eqnarray}
$\Delta x$ is the distance between the 2 motions, constant on a circular orbit,
and $F$ the component of the relative force projected along the mean motion.
$F$ may be expressed in terms of the circularly polarized metric perturbation 
$h$ which fits the circular motion of the planetary system 
\begin{eqnarray}
&&F\left( t\right)  = \frac{m\rho } {2\sqrt{2}} 
\left( h^{\prime \prime} e^{ 2i\Omega t} +
h^{*}\ ^{\prime \prime} e^{ -2i\Omega t} \right) \qquad
h\left( t\right)  =\frac{1}{\sqrt{2}}\left( h_{\rm 12} +
\frac{h_{\rm 22} -h_{\rm 11} }{2i}\right) 
\label{force_t}
\end{eqnarray}
The circular polarization $h$ is normalized so that it corresponds to
the same noise level as the two linear polarizations $h_{\rm 12}$ and 
$\frac{h_{\rm 22}-h_{\rm 11}}{2}$ in the case of an unpolarized
background. Polarizations are defined with respect to the plane 
of the orbit and the two motions chosen in the vicinity of $\theta=0$.

We now come to a Fourier representation of the force $F$ and metric $h$.
Accordingly, force and metric fluctuations are described by noise spectra 
$C_{FF}$ and $C_{hh}$ 
\begin{equation}
F\left[ \omega \right] =\int {\rm d}t\ F\left( t\right) \ e^{i\omega t}
\qquad
C_{FF}\left[ \omega \right] =\int {\rm d}t\ \left\langle F\left( t\right)
\cdot F\left( 0\right) \right\rangle \ e^{i\omega t}
\end{equation}
The dot specifies a symmetrical ordering when quantum fluctuations 
are considered \cite{Grishchuk77,Jaekel94}. 
The force (\ref{force_t}) is driven by gravitational waves through a 
frequency transposition due to the evolution of the quadrupole momentum 
at frequencies $\pm 2\Omega $ 
\begin{eqnarray}
&&F\left[ \omega \right] = -m\rho \frac{ 
\left( \omega + 2\Omega \right) ^{2} h\left[ \omega + 2\Omega \right] 
+ \left( \omega - 2\Omega \right) ^{2} h^{*}
\left[ \omega - 2\Omega \right] }{2\sqrt{2}}   
\end{eqnarray}
We are mainly interested in the long term cumulative effect of fluctuations, 
that is in the momentum perturbation $p_{t}$ integrated over an interaction 
time $t$ longer than the correlation time. 
The variance of $p_{t}$ is determined by the noise 
spectrum $C_{FF}\left[ 0\right]$ evaluated at zero frequency and 
expressed in terms of a momentum diffusion coefficient $D$ 
\begin{eqnarray}
&&p_{t} = \int_{0}^{t}{\rm d}s\ F\left( s\right) \qquad
\left\langle p_{t}^{2}\right\rangle = C_{FF}\left[ 0\right] \ t = 2Dt
\end{eqnarray}
The diffusion coefficient $D_{\rm gr}$ due to gravitational
waves is finally obtained as 
\begin{eqnarray}
&&2D_{\rm gr} = C_{FF}\left[ 0\right] =4m^{2}a^{2}C_{hh}\left[ 2\Omega
\right]   \qquad
a =\rho \Omega ^{2}  \label{diffCoeff}
\end{eqnarray}
$a$ is the acceleration on the circular orbit and $C_{hh}\left[ 2\Omega
\right] $ is the gravitational wave spectrum at frequency $2\Omega $ 
of the circular polarization $h$ coupled to the system. Note that a 
gravitational background of galactic origin is certainly not isotropic 
whereas an extragalactic background may probably be considered as isotropic. 
For simplicity, forthcoming discussions are phrased in terms of an 
unpolarized and isotropic background but the general case is recovered 
by coming back to equation (\ref{diffCoeff}).

The frequency of interest for the planetary motion of the Moon
is $\frac{2\Omega }{2\pi }\simeq 0.8\times 10^{-6}{\rm Hz}$.
Information on stochastic gravitational waves around this frequency 
may be deduced from studies devoted to the detectability of gravitational 
background by interferometers \cite{Schutz99,Ungarelli00}. We express
the spectrum $C_{hh}$ as a gravitational wave energy $k_{\rm B}T_{\rm gr}$ 
or, equivalently, as a number $n_{\rm gr}$ of gravitons per mode 
\begin{equation}
C_{hh}\left[ \omega \right] =\frac{16G}{5c^{5}}k_{\rm B}T_{\rm gr}\left[
\omega \right] =\frac{16G}{5c^{5}}\left( \frac{1}{2}+n_{\rm gr}\left[
\omega \right] \right) \hbar \omega 
\label{gravTemp}
\end{equation}
$k_{\rm B}$ is the Boltzmann constant and $T_{\rm gr}$ is an effective
temperature of the gravitational background.
Figure 1 of \cite{Schutz99} and equation (3.1) of \cite{Ungarelli00} lead
respectively to $C_{hh}\left[ 2\Omega \right] \simeq 10^{-34.5}{\rm Hz}^{-1}$
and $C_{hh}\left[ 2\Omega \right] \simeq 10^{-33}{\rm Hz}^{-1}$ for $\frac{%
2\Omega }{2\pi }\simeq 10^{-6}{\rm Hz}$. Taking as a conservative estimate $%
C_{hh}\simeq 10^{-34}{\rm Hz}^{-1}$, we obtain an effective temperature 
$T_{\rm gr}\approx 10^{41}{\rm K}$ or, equivalently, a graviton number per
mode $n_{\rm gr}\approx 2\times 10^{57}$. These numbers correspond to the
high temperature limit $k_{\rm B}T_{\rm gr}\gg \hbar \omega $.
Hence the gravitational noise is much larger than vacuum
fluctuations which correspond to the term $\frac{1}{2}$ in (\ref{gravTemp})
and lead to ultimate fluctuations of geodesic distances of the order of
Planck length \cite{Jaekel94}. 

The estimations of $C_{hh}$ used here
correspond to the confusion background of gravitational waves emitted by
binary systems in our galaxy. They rely on the laws of physics and
astrophysics as they are known in our local celestial environment. There
also exist more speculative predictions for gravitational backgrounds 
emitted by cosmic processes \cite{Schutz99,Ungarelli00,Grishchuk00,Maggiore00}. 
Depending on the parameters used in the models, these cosmic backgrounds 
may surpass the binary confusion background 
in the $\mu {\rm Hz}$ frequency range. 
Hence the latter can be considered as a minimum noise level in our
gravitational environment.

The diffusion coefficient may be written under the form of an Einstein 
fluctuation-dissipation relation, i.e. as the product of the effective 
temperature by a damping factor 
\begin{equation}
D_{\rm gr}=m\Gamma _{\rm gr}k_{\rm B}T_{\rm gr}\qquad \Gamma _{\rm gr}
=\frac{32Gma^{2}}{5c^{5}}  \label{EinsteinFDR}
\end{equation}
The damping rate $\Gamma _{\rm gr}$ is the inverse of the damping time
associated with the emission of gravitational waves by the planetary system 
\cite{Landau,Misner,Blanchet00}. It does not depend on temperature and is so
small, $\Gamma _{\rm gr}\approx 10^{-34}{\rm s}^{-1}$ for Moon, that it 
does not affect the classical motion on the orbit. Gravitational damping 
however has a noticeable effect in strongly bound binary systems such as 
millisecond pulsars \cite{Taylor92}. 

At this point it is worth comparing the effects of gravitational and 
electromagnetic scattering. Modelling the moving mass as a sphere which 
perfectly scatters thermal photons at temperature $T_{\rm em}$, 
the damping rate 
$\Gamma _{\rm em}$ associated with radiation pressure of these electromagnetic 
fluctuations is evaluated as \cite{Dalvit00} 
\begin{equation}
\Gamma _{\rm em}=\frac{4\pi ^{3} \hbar r^{2}}{45m} 
\left( \frac{k_{\rm B}T_{\rm em}}{\hbar c}\right) ^{4}
\end{equation}
The radius $r$ has been supposed to be large compared to the wavelength of the
photons. For the Moon, the temperature $T_{\rm em}=2.7{\rm K}$ of the cosmic
microwave background is already sufficient to produce a damping rate 
$\Gamma _{\rm em} \approx 2 \times 10^{-32}{\rm s}^{-1}$ which is 
$200$ times larger 
than $\Gamma _{\rm gr}$. Since $\Gamma _{\rm em}$ varies as $T_{\rm em}^{4}$ 
in agreement with Stefan-Boltzmann law, the damping due to solar radiation is 
more than $10^{10}$ times larger than $\Gamma _{\rm gr}$. 
In fact, the damping of the Moon, as it is revealed by the secular variation of
lunar rotation \cite{Bois96}, 
is mainly due to the interaction between Earth and 
Moon tides and it corresponds 
to a damping rate more than $10^{16}$ times larger 
than $\Gamma _{\rm gr}$. The role of gravitational scattering on damping 
of the Moon may thus be ignored but this is no longer the case for decoherence, 
as is shown in next paragraphs. 

In order to evaluate decoherence rates, we consider the effect of 
gravitational perturbation on the action difference $\Delta S_{t}$ 
after an interaction time $t$
\begin{eqnarray}
&&\Delta S_{t}= \int_{0}^{t}{\rm d}s F\left( s\right) \Delta x 
= p_{t} \Delta x  \qquad
\left\langle \Delta S_{t}^{2}\right\rangle = 
\left\langle p_{t}^{2}\right\rangle \Delta x^{2}
\end{eqnarray}
The decoherence between the two neighbouring trajectories is measured by the 
mean value $\left\langle \exp \frac{i\Delta S_{t}}{\hbar }\right\rangle $ of 
the exponentiated dephasing. Since $\Delta S_{t}$ is linearly driven 
by gravitational waves, it can be treated as a gaussian classical 
stochastic variable which leads to 
\begin{eqnarray}
&&\left\langle \exp \frac{i\Delta S_{t}}{\hbar }\right\rangle =
\exp \left( -\frac{\left\langle \Delta S_{t}^{2} \right\rangle }
{2\hbar ^{2}}\right) =\exp \left( -\Lambda _{\rm gr} \Delta x^{2} t\right)   
\nonumber \\
&&\Lambda _{\rm gr}=\frac{D_{\rm gr}}{\hbar ^{2}}
=\frac{32Gm^{2}a^{2}}{5c^{5}\hbar
^{2}} k_{\rm B}T_{\rm gr}  \label{Lambda}
\end{eqnarray}
Decoherence has been evaluated in the long term limit, where the force 
fluctuations may be approximated as a white noise characterized by a momentum 
diffusion coefficient $D_{\rm gr}$.
This expression can also be derived by evaluating the Feynman-Vernon 
influence functionals \cite{Feynman63}, often used in the study of decoherence 
\cite{Caldeira83}, at the limit of high temperature. Note that the
decoherence rate $\Lambda _{\rm gr}$ is proportional to the square of the
acceleration and, hence, would vanish if evaluated for an inertial motion. 
This has a simple interpretation through the Einstein formula (\ref
{EinsteinFDR}). The damping rate $\Gamma _{\rm gr}$ associated with the
emission of gravitational waves indeed vanishes for inertial motion while
the other factors entering the expression of $D_{\rm gr}$ do not
depend on the specific motion. 

Using the Einstein fluctuation-dissipation relation (\ref{EinsteinFDR}) and
the expression (\ref{Lambda}) of decoherence rates, the ratio between
gravitational and electromagnetic contributions to decoherence can be
rewritten 
\begin{equation}
\frac{\Lambda _{\rm gr}}{\Lambda _{\rm em}}=\frac{D_{\rm gr}}{D_{\rm em}}
=\frac{\Gamma _{\rm gr}}{\Gamma _{\rm em}}\frac{T_{\rm gr}}{T_{\rm em}}  
\end{equation}
For the motion of the Moon, 
the gravitational decoherence rate $\Lambda _{\rm gr}$ 
is found to be $10^{38}$ larger than the rate $\Lambda _{\rm em}$ corresponding 
to scattering of the cosmic photon background. It is still enormously larger 
than the effect corresponding to the scattering of solar photons. 
The same conclusion is reached for the comparison with the effect of tides.
The latter effect determines the damping of the main motion of Moon but 
its contribution to decoherence is overshadowed by gravitational scattering
owe to the very high value of the effective gravitational temperature. 
As a consequence of this high temperature, the decoherence time
is exceedingly short even for ultrasmall distances $\Delta x$. To fix ideas,
the gravitational decoherence rate $\Lambda _{\rm gr}\approx 10^{75}{\rm s}%
^{-1}{\rm m}^{-2}$ obtained for the motion of the Moon corresponds to a
decoherence time in the $10\mu{\rm s}$ range for a distance $\Delta x$ between 
two trajectories as small as Planck length.

The large value of gravitational temperature implies that decoherence should 
still be dominated by gravitational scattering for smaller 
size planetary systems. 
This can be discussed by writing the ratio between gravitational and
electromagnetic decoherence rates as a product of dimensionless factors 
\begin{eqnarray}
&&\frac{\Lambda _{\rm gr}}{\Lambda _{\rm em}} =
\frac{72}{\pi ^{3}} \frac{m^{2}}{m_{\rm P}^{2}} \frac{\rho ^{2}}{r^{2}} 
\left( \frac {\hbar \Omega } {k_{\rm B}T_{\rm em}} \right) ^{4}
\frac{T _{\rm gr}}{T _{\rm em}} 
\label{ratioLambda}
\end{eqnarray}
The factor $\frac{m^{2}}{m_{\rm P}^{2}}$ is clearly reminiscent of
the simple scaling arguments presented in the Introduction, illustrating
the role of Planck mass as a reference on the borderland between
microscopic and macroscopic masses. However the whole
formula shows that these scaling arguments are not sufficient for obtaining
reliable quantitative predictions. The factor $\frac{\rho ^{2}}{r^{2}}
$ is a geometrical factor depending on the radius $\rho $ of the orbit and
the radius $r$ of the orbiter. Then the ratio (\ref{ratioLambda}) also 
depends on the inverse fourth power of the photon number 
$\frac{k_{\rm B}T_{\rm em}}{\hbar \Omega }$ per mode at the orbital frequency
$\Omega $ and on the already discussed ratio between temperatures of
graviton background and photon background. Clearly the last three 
dimensionless factors appearing in (\ref{ratioLambda}) have
no relation with the scaling arguments of the Introduction but 
have to be known for a quantitative comparison of gravitational and
electromagnetic contributions. If we consider as an example a man made
gravitationally bound planetary system consisting of two spheres having an
ordinary metallic density, we obtain comparable decoherence rates for
gravitational scattering and electromagnetic scattering of cosmic microwave 
background in the case of a compact geometry with masses of the order of 
$10^{3}{\rm kg}$. 

We have shown in this letter that the scattering of gravitational waves in
our celestial environment is a dominant cause of decoherence for planetary
motions such as the motion of the Moon around the Earth. Due to the very large
effective temperature of the gravitational wave background, this mechanism
leads to a decoherence by far more efficient 
than the other fluctuation mechanisms 
though its contribution to damping of the mean motion can be neglected. 
As far as the theory of measurement is concerned, this implies that 
the ultimate fluctuations of the motion of Moon are determined by 
the same classical gravitation theory which also explains its mean
motion. Precisely these fluctuations are determined by the classical 
gravitational waves present in our local celestial environment. 

\acknowledgments
P.A.M.N. wishes to thank CAPES, CNPq,
FAPERJ, PRONEX, COFECUB, ENS and MENRT for their financial support which
made possible his stays in Paris during which this work was performed.


\begin{thebibliography}{0}

\bibitem{Zeh70}  
\Name{Zeh H.D.} \REVIEW{Found. Phys.}{1}{1970}{69}.

\bibitem{Dekker77}  
\Name{Dekker H.} \REVIEW{Phys. Rev.}{A16}{1977}{2126}.

\bibitem{Zurek81}  
\Name{Zurek A.J.} \REVIEW{Phys. Rev.}{D24}{1981}{1516};
\REVIEW{Phys. Rev.}{D26}{1982}{1862}.

\bibitem{Joos85}  
\Name{Joos E. \and Zeh H.D.} \REVIEW{Z. Phys.}{B59}{1985}{223}.

\bibitem{Brune96}  
\Name{Brune M., Hagley E., Dreyer J. \etal}
\REVIEW{Phys. Rev. Lett.}{77}{1996}{4887}; 
\Name{Davidovich L., Brune M., Raimond J.M. \and Haroche S.} 
\REVIEW{Phys. Rev.}{A53}{1996}{1295}.

\bibitem{Wheeler57}  
\Name{Wheeler J.A.} \REVIEW{Ann. Physics}{2}{1957}{604}.

\bibitem{DeWitt62}  
\Name{DeWitt B.S.} \Book{Gravitation, An Introduction to
Current Research} \Editor{L.Witten} \Publ{Wiley} \Year{1962} \Page{266}.

\bibitem{Karolyhazy66}  
\Name{Karolyhazy F.} \REVIEW{Nuovo Cim.}{42A}{1966}{390}.

\bibitem{Diosi89} 
\Name{Diosi L.} \REVIEW{Phys. Rev.}{A40}{1989}{1165}.

\bibitem{Penrose96} 
\Name{Penrose R.} \REVIEW{Gen. Rel. Grav.}{28}{1996}{581}.

\bibitem{Hils90}  
\Name{Hils D., Bender P.L. \and Webbink R.F.}
\REVIEW{Astrophys. J.}{360}{1990}{75}.

\bibitem{Giazotto97} 
\Name{Giazotto A., Bonazzola S. \and Gourgoulhon E.}
\REVIEW{Phys. Rev.}{D55}{1997}{2014}.

\bibitem{Landau}  
\Name{Landau L.D. \and Lifshitz E.M.} \Book{The Classical Theory
of Fields} \Publ{Butterworth-Heinemann} \Year{1998} § 110.

\bibitem{Misner}  
\Name{Misner C.W., Thorne K.S. \and Wheeler J.A.}
\Book{Gravitation} \Publ{Freeman} \Year{1973} § 37.

\bibitem{Blanchet00}  
\Name{Blanchet L., Kopeikin S. \and Sch\"{a}fer G.}
\REVIEW{arXiv}{}{2000}{gr-qc/0008074}.

\bibitem{Grishchuk77}  
\Name{Grishchuk L.P.} 
\REVIEW{Usp. Fiz. Nauk}{121}{1977}{629}
[\REVIEW{Sov. Phys. Usp.}{20}{1977}{319}]; 
\REVIEW{Usp. Fiz. Nauk}{156}{1988}{297} 
[\REVIEW{Sov. Phys. Usp.}{31}{1988}{940}].

\bibitem{Jaekel94}  
\Name{Jaekel M.T. \and Reynaud S.} 
\REVIEW{Phys. Lett.}{A185}{1994}{143}; 
\REVIEW{Quant. Semicl. Optics}{7}{1995}{639};
\REVIEW{Annalen der Physik}{4}{1995}{68}.

\bibitem{Schutz99}  
\Name{Schutz B.} \REVIEW{Class. Quant. Grav.}{16}{1999}{A131}.

\bibitem{Ungarelli00}  
\Name{Ungarelli C. \and Vecchio A.} 
\REVIEW{arXiv}{}{2000}{gr-qc/0003021}.

\bibitem{Grishchuk00}  
\Name{Grishchuk L.P., Lipunov V.M., Postnov K.A.
\etal} \REVIEW{arXiv}{}{2000}{astro-ph/0008481}.

\bibitem{Maggiore00}  
\Name{Maggiore M.} \REVIEW{Phys. Rep.}{331}{2000}{283}.

\bibitem{Taylor92}  
\Name{Taylor J.H., Wolszczan A., Damour T.
\and Weisberg J.M.} \REVIEW{Nature}{355}{1992}{132}.

\bibitem{Dalvit00} 
\Name{Dalvit D. \and Maia Neto P.A.} 
\REVIEW{Phys. Rev. Lett.}{84}{2000}{798}; 
\Name{Maia Neto P.A. \and Dalvit D.} 
\REVIEW{Phys. Rev.}{A62}{2000}{042103}.

\bibitem{Bois96}  
\Name{Bois E., Boudin F. \and Journet A.}
\REVIEW{Astron. Astrophys.}{314}{1996}{989}.

\bibitem{Feynman63}  
\Name{Feynman R.P. \and Vernon F.L.} 
\REVIEW{Ann. Physics}{24}{1969}{118}.

\bibitem{Caldeira83}  
\Name{Caldeira A.O. \and Leggett A.J.} 
\REVIEW{Physica}{121A}{1983}{587}; 
\REVIEW{Phys. Rev.}{A31}{1985}{1059}.
\end{thebibliography}
\end{document}